\documentclass[12pt,a4paper]{article}
\usepackage{graphicx}
\usepackage{amsfonts} 
\usepackage{amssymb}
\usepackage{amsthm}
\usepackage{amsmath}
\usepackage[a4paper,margin=4cm]{geometry}
\usepackage{lastpage}
\usepackage{fancyhdr}
\usepackage{cite}

\newcommand{\shorttitlestr}{Compact Set Packing}
\newcommand{\authorstr}{Huairui Chu} 

\newtheorem{theorem}{Theorem}
\newtheorem{lemma}{Lemma}
\newtheorem{definition}{Definition}

\begin{document}
\author{Huairui Chu}
\title{A Tight Lower Bound for Compact Set Packing}
\date{\today}
\maketitle

\pagestyle{fancy}
\fancyhf{}
\rhead{\shorttitlestr}
\lhead{\authorstr}
\rfoot{Page \thepage\ of \pageref{LastPage}}
\renewcommand{\headrulewidth}{1pt}


This note is devoted to show a simple proof of a tight lower bound of the parameterized compact set packing problem, based on ETH. To see a more comprehensive study on this problem, we refer to \cite{psp}.

Notice that according to the well-known sparsification lemma \cite{sparsificationlemma}, ETH implies that sparse $3$-SAT can't be solved in time $2^{o(n)}$. ``Sparse'' means the number of clauses $m$ is at most linear in $n$, i.e. $m=O(n)$.
\begin{definition}
    \textbf{(Linear) Parameterized Compact Set Packing} \\
    Instance: Universe $\mathcal{U}$, set family $\mathcal{S}$, parameter $r$. $|U|=f(r)\Theta(\log|\mathcal{S}|)$. \\
    Question: Is there $r$ pair-wise disjoint sets from $\mathcal{S}$?
\end{definition}

\section{A Tight Lower Bound from ETH}

A brute-force algorithm for this problem runs in time $|\mathcal{S}|^{O(r)}|U|^{O(1)}$, we provide a lower bound of the form $|\mathcal{S}|^{\Omega(r)}$.

\begin{theorem}\label{main}
    There is no algorithm running in time $|\mathcal{S}|^{o(r)}$ for Compact Set Packing, whose instances satisfy $|U|=r^3\Theta(\log |\mathcal{S}|)$.
\end{theorem}

To prove this theorem, we devise a reduction from sparse $3$-SAT to compact set packing.

\begin{lemma}\label{reduction}
    For all $r$ a computable function of $n$, there exists a reduction such that, given a sparse $3$-SAT instance $\phi$, a Compact Set Packing instance $I$ is created such that $\phi$ is satisfiable if and only if $I$ has an $r$-set packing. The reduction running in time $2^{O(n/r)}poly(n,r)$, and $I$ has $\Theta(nr^2)$-size universe and $\Theta(r2^{O(n/r)})$ sets.
\end{lemma}

\begin{proof}
    Given a $3$-SAT instance $\phi$, in which there are $n$ variables and $m=O(n)$ clauses. Let $r$ be an integer. We create a Set Packing instance in the following.\\
    Partition the clauses into $r$ disjoint sets $C_1,C_2,...,C_r$ such that $||C_i|-\frac{m}{r}|\leq 1$ for all $i\in [r]$. Let $v(C_i)$ be the variables that appear in any clause in $C_i$. For each $i\in [r]$, let $A_i$ be all the partial assignment from $v(C_i)$ to $\{0,1\}$ such that each clause in $C_i$ is satisfied. There are at most $2^{|v(C_i)|}=2^{O(n/r)}$ such assignments.\\
    For each variable $x$, let $G_x=(U_x\cup V_x,E_x)$ be a complete bipartite graph where $U_x = \{u^x_1,u^x_2,...,u^x_r\},V_x = \{v^x_1,v^x_2,...,v^x_r\}$. For each $i\in [r]$ and each partial assignment $a\in A_i$, we add a set $S_a$. For each $x\in domain(a)$, if $a(x)=0$, we add $(u^x_i,v^x_j)$ for all $j\in [r]$ to $S_a$, if $a(x)=1$, we add $(u^x_j,v^x_i)$ for all $j\in [r]$ to $S_a$. Let $\mathcal{S}_i$ be the sets $\{S_a|a\in A_i\}$. \\
    Each $\mathcal{S}_i$ contains at most $2^{O(n/r)}$ sets, and $O(nr)$ elements. For each $\mathcal{S}_i$ we create an ISS of universe size $O(n/r)$ containing $|\mathcal{S}_i|$ pairwise disjoint sets. Join each set $S_a\in \mathcal{S}_i$ to a unique set in the ISS. To be specific, let $\sigma_i$ be an injection from $\mathcal{S}_i$ to the sets of the ISS, we create a new set family $\mathcal{S}'_i=\{S_a\cup \sigma(S_a)|S_a\in \mathcal{S}_i\}$.\\
    Let $X$ be the variable set of $\phi$. The set packing instance we created is $(\mathcal{U},\mathcal{S})$ where $\mathcal{S}=\cup_{i\in [r]}\mathcal{S}'_i$ and $\mathcal{U}=\cup_{x\in X}E_{x}$. The parameter is set to be $r$. It's not hard to see the correctness. The universe size is $\Theta(nr^2)$, the set family size is $O(r2^{O(n/r)})$. 
    
    To make it a compact instance, we can add sets such that the set family size is $\Theta(r2^{O(n/r)})$. One possible way for adding such sets is to consider a set $\mathcal{D}$ of $\Theta(n\log r/r)$ dull elements to the universe. We just add $\mathcal{U}\cup D$ as a new set, for all $D\subseteq \mathcal{D}$. So that the final instance is $(\mathcal{U'},\mathcal{S}'
    )$ where $\mathcal{U}'=\mathcal{U}\cup \mathcal{D}$, $\mathcal{S}'=\mathcal{S}\cup \{\mathcal{U}\cup D|\forall D\subseteq \mathcal{D}\}$.
\end{proof}

Based on Lemma \ref{reduction}, if Compact Set Packing admits $|\mathcal{S}|^{o(r)}$ time algorithm, there will be a $r^{o(r)}2^{o(n)}+2^{O(n/r)}poly(n,r)$ algorithm for (linear) $3$-SAT for all $r$. Set $r$ to $\lceil\log n\rceil$, a contradiction to ETH. This indicates that there is no algorithm running in time $|S|^{o(r)}$ for Compact Set Packing instances where $|U|=r^3\Theta(\log |\mathcal{S}|)$. Thus Theorem \ref{main} is proved.

\bibliographystyle{abbrv}
\bibliography{bib_file.bib}

\end{document}